\documentclass[a4paper,twocolumn,english,prl,showpacs,superscriptaddress]{revtex4}
\usepackage{mathpazo}

\usepackage[T1]{fontenc}
\usepackage[latin9]{inputenc}
\usepackage{fancyhdr}
\usepackage{babel}
\usepackage{amsmath}
\usepackage{amssymb}
\usepackage{graphicx}
\usepackage[unicode=true,pdfusetitle,
 bookmarks=true,bookmarksnumbered=false,bookmarksopen=false,
 breaklinks=false,pdfborder={0 0 1},backref=false,colorlinks=false]
 {hyperref}
\usepackage{breakurl}

\makeatletter

\@ifundefined{textcolor}{}
{%
 \definecolor{BLACK}{gray}{0}
 \definecolor{WHITE}{gray}{1}
 \definecolor{RED}{rgb}{1,0,0}
 \definecolor{GREEN}{rgb}{0,1,0}
 \definecolor{BLUE}{rgb}{0,0,1}
 \definecolor{CYAN}{cmyk}{1,0,0,0}
 \definecolor{MAGENTA}{cmyk}{0,1,0,0}
 \definecolor{YELLOW}{cmyk}{0,0,1,0}
}

\usepackage{fancyhdr}

\usepackage{babel}
\usepackage{breakurl}

\makeatother

\begin{document}

\title{All-quantum teleportation}

\author{N. G. de Almeida, }

\affiliation{Instituto de Física, Universidade Federal de Goiás, 74.001-970, Goiânia
- GO, Brazil}
\begin{abstract}
I propose to replace the dual classical and nonlocal channels used
for teleporting unknown quantum states in the original protocol (OP)
{[}Bennett, C. H.,\emph{ et al.} Phys. Rev. Lett. \textbf{70} 1895
(1993){]} by either (i) one single quantum channel or (ii) two nonlocal
channels in order to turn the OP into an all-quantum teleportation
(AQT) protocol. Ideally, N runs of single channel AQT can be achieved
with a single Einstein-Podolsky-Rosen (EPR) pair, in contrast with
the OP, which consumes N EPR pairs. In the two nonlocal channels proposal,
Alice uses the superdense coding technique to send Bob her result,
which makes AQT more economical than OP. 
\end{abstract}
\maketitle
\textit{Introduction. }\textit{\emph{Quantum teleportation was suggested
by Bennett }}\textit{et al}\textit{\emph{. \cite{Bennet93} and is
recognized nowadays as one of the cornerstones of quantum computation
and quantum communication \cite{Nielsen00,Gottesman99,Lo98}. Two
important features characterizing this phenomenon is the transfer
of information between noninteracting systems by means of a dual quantum
and classical channels. Indeed, in the very tittle of the pioneering
work Ref. \cite{Bennet93}, which henceforth will be called the }}\textit{original
protocol }\textit{\emph{(OP), the classical resource is highlighted,
and in the abstract the authors says \textquotedbl{}Alice makes a
joint measurement on her EPR particle and the unknown quantum system,
and sends Bob the classical result of this measurement.\textquotedbl{}
\cite{Bennet93}. This classical result consists of two classical
bits out of the combination $(00,01,10,11)$. }}

Soon after the work by Bennett \emph{et al}., several other theoretical
schemes for teleportation differing from the OP were proposed. As
for example, in Ref.\cite{Vaidman94} it was proposed a ``cross measurement\textquotedblright{}
method to achieve \emph{a two-way teleportation}, and in Ref.\cite{moussa97}
the author shows how to implement teleportation \emph{with identity
interchange of quantum states}. Yet, in Ref. \cite{deAlmeida98},
the authors used Greenberger\textendash Horne\textendash Zeilinger
(GHZ) states as the nonlocal channel instead of the Einstein\textendash Podolsky\textendash Rosen
(EPR) states used in the OP, and in Ref.\cite{Karlson98}, also using
a GHZ channel, the authors show how to accomplish teleportation controlled
by a third party. In Ref.\cite{Agrawal06} the authors show how to
achieve perfect teleportation and superdense code using W states as
the nonlocal channel. In Refs.\cite{Cardoso08,Cardoso09} the concept
of controlled partial teleportation (CPT) was introduced\textbf{.
}Recently, a protocol to send information in a secure way inspired
by the OP was proposed, which used a classical correlated channel
\cite{Costa16}.

\textit{\emph{Here I propose a modification in the OP to turn it into
an all-quantum teleportation (AQT) protocol by either (i) abolishing
the classical channel and using one single quantum channel, or (ii)
replacing the classical channel by a quantum one and using two nonlocal
channels. In (i), a joint quantum nondemolition (QND) measurement
}}\cite{Guo01}\textit{\emph{ is used, such that the particle whose
unknown state is in Alice's hands has a twofold usefulness: it is
used both for Alice sending Bob her joint QND measurement result and
also for reestablishing the nonlocal channel when Bob sends back Alice
one particle to re-initiate the AQT protocol. To accomplish this,
Alice and Bob must be able to perform nonlocal quantum nondemolition
measurement and to store and send the information encoded in their
particles. Next, I show how this can be theoretically accomplished.}}

\emph{Nonlocal channels for all-quantum teleportation}. As in the
OP, Alice and Bob share a nonlocal channel, which can be any of the
four states composing the complete and orthonormal Bell basis for
particles $A$ and $B$ $\left\{ \left|\psi^{\pm}\right\rangle _{AB},\left|\phi^{\pm}\right\rangle _{AB}\right\} $,
where

\begin{equation}
\left|\psi^{\pm}\right\rangle _{AB}=\sqrt{\frac{1}{2}}\left(\left|01\right\rangle _{AB}\pm\left|10\right\rangle _{AB}\right),\label{eq:1}
\end{equation}
\begin{equation}
\left|\phi^{\pm}\right\rangle _{AB}=\sqrt{\frac{1}{2}}\left(\left|00\right\rangle _{AB}\pm\left|11\right\rangle _{AB}\right).\label{eq:2}
\end{equation}
When Alice receives from Charles the state to be teleported to Bob,
which can be cast in the form $\left|\chi\right\rangle _{C}=\alpha\left|0\right\rangle _{C}+\beta\left|1\right\rangle _{C}$,
with $\left|\alpha\right|^{2}+\left|\beta\right|^{2}=1$, she performs
a nonlocal or joint measurement on particles $A$ and $C$. Before
Alice's measurement, there will be four possibilities to the complete
state of the three particles, depending on which of the four nonlocal
channels, Eqs.(\ref{eq:1}) and (\ref{eq:2}), Alice and Bob are sharing.
After knowing that Alice's measurement was one of these four $\left|\psi^{\pm}\right\rangle $
or $\left|\phi^{\pm}\right\rangle $ states, Bob applies one of the
four local unitary operation to complete teleportation: $\mathbf{1},\sigma_{z},\sigma_{x},\sigma_{x}\sigma_{z}$,
depending on the nonlocal channel used. Explicitly, if the nonlocal
channel is $\left|\psi^{-}\right\rangle _{AB}$, the complete state
of the three particles $A$, $B$ and $C$ will be $\left|\psi\right\rangle _{ABC}=\left|\psi^{-}\right\rangle _{AB}\left|\chi\right\rangle _{C}$,
which can be written in the following way 
\begin{align}
\left|\psi\right\rangle _{ABC} & =\frac{1}{2}\left|\psi^{+}\right\rangle _{AC}\left(\alpha\left|0\right\rangle _{B}-\beta\left|1\right\rangle _{B}\right)\nonumber \\
 & +\frac{1}{2}\left|\psi^{-}\right\rangle _{AC}\left(\alpha\left|0\right\rangle _{B}+\beta\left|1\right\rangle _{B}\right)\nonumber \\
 & +\frac{1}{2}\left|\phi^{+}\right\rangle _{AC}\left(-\alpha\left|1\right\rangle _{B}+\beta\left|0\right\rangle _{B}\right)\nonumber \\
 & -\frac{1}{2}\left|\phi^{-}\right\rangle _{AC}\left(\alpha\left|1\right\rangle _{B}+\beta\left|0\right\rangle _{B}\right),\label{Bell psi minus}
\end{align}
where now the subscripts in Bell states is for $A$ and $C$. After
the Alice's joint measurement, particles state $A$ and $C$ gets
entangled in one of the four Bell states, and the local unitary operation
Bob must apply to convert his particle state into a perfect copy of
$\left|\chi\right\rangle _{C}$ is either $\mathbf{1},$ or $\sigma_{z}$,
or $\sigma_{x}$, or $\sigma_{x}\sigma_{z}$ if Alice's result was
either $\left|\psi^{-}\right\rangle _{AC}$ or $\left|\psi^{+}\right\rangle _{AC}$
or $\left|\phi^{-}\right\rangle _{AC}$, or $\left|\phi^{+}\right\rangle _{AC}$
, respectively.

If the nonlocal channel is $\left|\psi^{+}\right\rangle $, the complete
state can be written as 
\begin{align}
\left|\psi\right\rangle _{ABC} & =\frac{1}{2}\left|\psi^{+}\right\rangle _{AC}\left(\alpha\left|0\right\rangle _{B}+\beta\left|1\right\rangle _{B}\right)\nonumber \\
 & -\frac{1}{2}\left|\psi^{-}\right\rangle _{AC}\left(\alpha\left|0\right\rangle _{B}-\beta\left|1\right\rangle _{B}\right)\nonumber \\
 & +\frac{1}{2}\left|\phi^{+}\right\rangle _{AC}\left(-\alpha\left|1\right\rangle _{B}+\beta\left|0\right\rangle _{B}\right)\nonumber \\
 & -\frac{1}{2}\left|\phi^{-}\right\rangle _{AC}\left(\alpha\left|1\right\rangle _{B}+\beta\left|0\right\rangle _{B}\right),\label{eq:Bell psi plus}
\end{align}
and the local unitary operation that Bob must apply is either $\mathbf{1},$
or $\sigma_{z}$, or $\sigma_{x}$, or $\sigma_{x}\sigma_{z}$ if
Alice's result was either $\left|\psi^{+}\right\rangle _{AC}$ or
$\left|\psi^{-}\right\rangle _{AC}$ or $\left|\phi^{-}\right\rangle _{AC}$,
or $\left|\phi^{+}\right\rangle _{AC}$ , respectively.

If If the nonlocal channel is $\left|\phi^{-}\right\rangle $, the
complete $\left|\psi\right\rangle _{ABC}$ can be written as 
\begin{align}
\left|\psi\right\rangle _{ABC} & =\frac{1}{2}\left|\psi^{+}\right\rangle _{AC}\left(-\alpha\left|1\right\rangle _{B}+\beta\left|0\right\rangle _{B}\right)\nonumber \\
 & -\frac{1}{2}\left|\psi^{-}\right\rangle _{AC}\left(\alpha\left|1\right\rangle _{B}+\beta\left|0\right\rangle _{B}\right)\nonumber \\
 & +\frac{1}{2}\left|\phi^{+}\right\rangle _{AC}\left(\alpha\left|0\right\rangle _{B}-\beta\left|1\right\rangle _{B}\right)\nonumber \\
 & +\frac{1}{2}\left|\phi^{-}\right\rangle _{AC}\left(\alpha\left|0\right\rangle _{B}+\beta\left|1\right\rangle _{B}\right),\label{eq:Bell phi minus}
\end{align}
and the local unitary operation that Bob must apply is either $\mathbf{1},$
or $\sigma_{z}$, or $\sigma_{x}$, or $\sigma_{x}\sigma_{z}$ if
Alice's result was either $\left|\phi^{-}\right\rangle _{AC}$ or
$\left|\phi^{+}\right\rangle _{AC}$ or $\left|\psi^{-}\right\rangle _{AC}$,
or $\left|\psi^{+}\right\rangle _{AC}$ , respectively.

Finally, if If the nonlocal channel is $\left|\phi^{+}\right\rangle $
, $\left|\psi\right\rangle _{ABC}$ can be written as 
\begin{align}
\left|\psi\right\rangle _{ABC} & =\frac{1}{2}\left|\psi^{+}\right\rangle _{AC}\left(\alpha\left|1\right\rangle _{B}+\beta\left|0\right\rangle _{B}\right)\nonumber \\
 & -\frac{1}{2}\left|\psi^{-}\right\rangle _{AC}\left(\alpha\left|1\right\rangle _{B}-\beta\left|0\right\rangle _{B}\right)\nonumber \\
 & +\frac{1}{2}\left|\phi^{+}\right\rangle _{AC}\left(\alpha\left|0\right\rangle _{B}+\beta\left|1\right\rangle _{B}\right)\nonumber \\
 & +\frac{1}{2}\left|\phi^{-}\right\rangle _{AC}\left(\alpha\left|0\right\rangle _{B}-\beta\left|1\right\rangle _{B}\right),\label{eq:Bell phi plus}
\end{align}
and the local unitary operation Bob must apply is either $\mathbf{1},$
or $\sigma_{z}$, or $\sigma_{x}$, or $\sigma_{x}\sigma_{z}$ if
Alice's result was either $\left|\phi^{+}\right\rangle _{AC}$ or
$\left|\phi^{-}\right\rangle _{AC}$ or $\left|\psi^{+}\right\rangle _{AC}$,
or $\left|\psi^{-}\right\rangle _{AC}$ , respectively. In the original
protocol, right after Alice did the measurement, she sends two classical
bits out of $\left\{ 00,01,10,11\right\} $ representing one of the
four possible results $\left|\psi^{\pm}\right\rangle $ and $\left|\phi^{\pm}\right\rangle $
above. In AQT, Bob becomes aware of Alice's measurement when he receives
not the classical bits, but the quantum ones, either $\left|\psi^{\pm}\right\rangle $
or $\left|\phi^{\pm}\right\rangle $, sent by Alice. This can be done
in two different ways: (i) in the single channel AQT, after Alice's
performs a \emph{joint QND measurement}, which can be accomplished
using two ancillary states of additional particles $D$ and $E$ interacting
with particles $A$ and $C$ according to the circuit in Fig.1. 
\begin{figure}[ptb]
\centering{}\includegraphics[width=1\columnwidth]{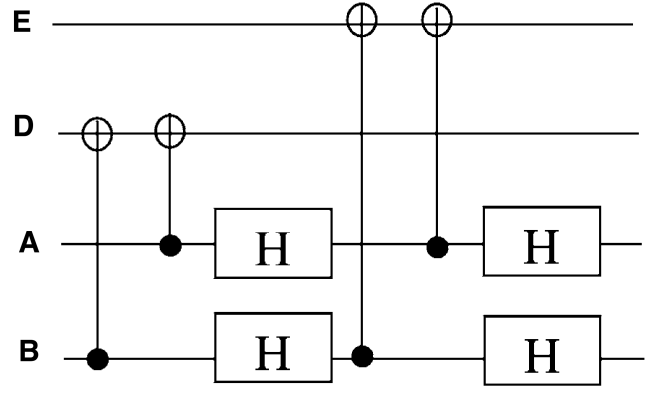}\caption{Circuit network to implement the QND measurement needed to accomplish
single channel AQT. The ancilla particle state $D$ flips controlled
by the particle states $B$ and $A$, respectively. After that, a
Hadarmard gate is applied on particles $A$ and $B$. Finally, the
ancilla particle state $E$ flips controlled by particle states $B$
and $A$, respectively.}
\end{figure}

After the QND joint measurement, Alice sends Bob the two particles
$A$ and $C$ whose entangled state contains the message on the unitary
operation Bob must apply to complete AQT. (ii) In the two nonlocal
channels, Alice uses one channel to teleport to Bob the state of particle
$C$ and the second channel two send quantum information containing
the message on the unitary operation Bob must apply to complete AQT.
Next, I discuss this two ways for AQT in more details.

\emph{Single channel all-quantum teleportation}. Let the nonlocal
channel initially shared by Alice and Bob be $\left|\psi^{-}\right\rangle _{AB}$,
such that the complete state before Alice's measurement is given by
Eq.(\ref{Bell psi minus}).

The sequence of unitary operations described in Fig. 1 are the following:
The ancilla particle state $D$ flips controlled by the particle states
$B$ and $A$, respectively. After that, a Hadarmard gate is applied
on particles $A$ and $B$. Finally, the ancilla particle state $E$
flips controlled by particle states $B$ and $A$, respectively. Starting
from Eq.(\ref{Bell psi minus}), it is straightforward to calculate
the complete state after interacting with ancillas $E$ and $D$ as
given by the circuit of Fig.1: 
\begin{align}
\left|\psi\right\rangle _{ABC} & \rightarrow\frac{1}{2}\left|\psi^{+}\right\rangle _{AC}\left(\alpha\left|0\right\rangle _{B}-\beta\left|1\right\rangle _{B}\right)\left|1\right\rangle _{D}\left|0\right\rangle _{E}\nonumber \\
 & +\frac{1}{2}\left|\psi^{-}\right\rangle _{AC}\left(\alpha\left|0\right\rangle _{B}+\beta\left|1\right\rangle _{B}\right)\left|1\right\rangle _{D}\left|1\right\rangle _{E}\nonumber \\
 & +\frac{1}{2}\left|\phi^{+}\right\rangle _{AC}\left(-\alpha\left|1\right\rangle _{B}+\beta\left|0\right\rangle _{B}\right)\left|0\right\rangle _{D}\left|0\right\rangle _{E}\nonumber \\
 & -\frac{1}{2}\left|\phi^{-}\right\rangle _{AC}\left(\alpha\left|1\right\rangle _{B}+\beta\left|0\right\rangle _{B}\right)\left|0\right\rangle _{D}\left|1\right\rangle _{E}.\label{Bell psi minus-1}
\end{align}
From Eq.(\ref{Bell psi minus-1}) we can clearly see that Alice's
measurement on ancilla states $D$ and $E$ allows her to discern
in which of the four states of Bell basis particles $A$ and $C$
collapse. For example, assume the Alice's result was $\left|1\right\rangle _{D}\left|0\right\rangle _{E}$.
Then, the collapsed state is the tensorial product $\left|\psi\right\rangle _{ABC}\rightarrow\left|\psi^{+}\right\rangle _{AC}\left(\alpha\left|0\right\rangle _{B}-\beta\left|1\right\rangle _{B}\right)\left|1\right\rangle _{D}\left|0\right\rangle _{E}$,
and Bob is left with state $\alpha\left|0\right\rangle _{B}-\beta\left|1\right\rangle _{B}$.
Alice thus sends Bob particles $A$ and $C$ in state $\left|\psi^{+}\right\rangle _{AC}$
altogether, and Bob applies the same operations needed to make the
QND measurement described in Fig.1, with the label $B$ replaced by
$C$ and keeping the same labels $D$ and $E$ to his ancilla states.
Bob's result is, of course, $\left|\psi^{+}\right\rangle _{AC}\left|1\right\rangle _{D}\left|0\right\rangle _{E}$.
Thus, after measuring the ancilla in states $\left|1\right\rangle _{D}\left|0\right\rangle _{E}$
, Bob applies the local unitary operation $\sigma_{z}$ on the particle
$B$ to convert its state in a perfect replica of particle $C$ and
completing one run of single-channel AQT.

To complete subsequent runs of AQT, Bob and Alice have two alternatives
which, of course, needs to be previously combined between them: (i)
According to Fig.2, Bob applies a local unitary operation $U_{A}$
to reestablish the original nonlocal state, which, in the case exemplified
above, can be $\sigma_{z}^{(A)}\left|\psi^{+}\right\rangle _{AC}\rightarrow\left|\psi^{-}\right\rangle _{AC}$,
and sends back particle $A$ to Alice, keeping particle $C$ with
him relabeling $C\rightarrow B$. Once reestablished the nonlocal
channel $\left|\psi^{+}\right\rangle _{AC}\rightarrow\left|\psi^{-}\right\rangle _{AB}$,
Alice then repeats the nonlocal QND measurement on particles $A$
and $C$ and the protocol continues unchanged. (ii) Bob sends Alice
exactly the state he has measured, in this example $\left|\psi^{+}\right\rangle _{AC}$.
Alice, knowing that Bob has just send back particle $A$, makes a
nonlocal measurement having in mind the nonlocal channel as given
by Eq.(\ref{eq:Bell psi plus}). Alice performs the nonlocal QND measurement
and sends her result to Bob who, after performing a nonlocal QND measurement,
applies one of the following local operation to complete the second
run of AQT: $\mathbf{1},$ $\sigma_{z}$, $\sigma_{x}$, or $\sigma_{x}\sigma_{z}$
if Alice's result was either $\left|\psi^{+}\right\rangle _{AC}$
or $\left|\psi^{-}\right\rangle _{AC}$ or $\left|\phi^{-}\right\rangle _{AC}$,
or $\left|\phi^{+}\right\rangle _{AC}$, respectively - see Eq.(\ref{eq:Bell psi plus}).
A third run will follow a similar procedure, with the subsequent nonlocal
channel being one of those four states Fig.(\ref{Bell psi minus})-(\ref{eq:Bell phi plus}).
Fig.2 displays the circuit for the one single channel AQT protocol
using approach (i). 
\begin{figure}[ptb]
\centering{}\includegraphics[width=1\columnwidth]{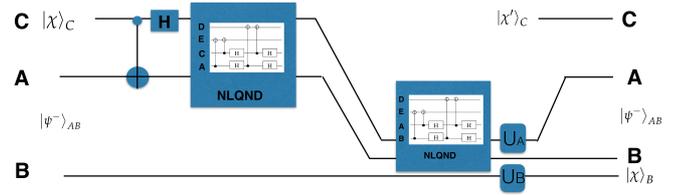}\caption{Circuit network to implement single channel all-quantum teleportation.
Alice (A) and Bob (B) share one single nonlocal channel $\left|\psi^{-}\right\rangle _{AB}$.
Charles (C) delivers Alice the state $\left|\chi\right\rangle _{C}$
to be teleported to Bob. After the CNOT and Hadamard gates, Alice
performs a joint QND measurement on particles $A$ and $C$ using
two ancillas $D$ and $E$ and sends Bob both particles. Bob performs
another joint QND measurement on this both particles to know the local
unitary operation $U_{B}$ he must apply to his particle $B$ to convert
its state into a perfect copy of particle's state $C$ . After that,
he applies a local unitary operation $U_{A}$ on the particle of the
pair $A$ and $C$ and sends back Alice particle C, restoring the
initial nonlocal channel $\left|\psi^{-}\right\rangle _{AB}$. The
apparatus is then prompt to another run of AQT.}
\end{figure}

We note that the second approach will be useful, and even necessary,
if any of the unitary local operations is unavailable. This is the
case, for example, if we are working with Fock states, for which no
Hadamard gate exists. Although conceptually simple, AQT may pose some
challenge in its accomplishment, mainly due to the maintenance of
such a fragile resource as entanglement. Notwithstanding, it would
be interesting to known how many runs of single channel AQT can be
achieved based on existing technology.

\emph{Two nonlocal channels all-quantum teleportation. }Let us now
show that AQT can be achieved replacing the classical channel by a
nonlocal channel and thus using two nonlocal channels instead of a
dual classical and nonlocal channels as in Ref.\cite{Bennet93}. Indeed,
one of the two channels can be used by Alice to communicate, e.g.,
via superdense coding \cite{Nielsen00,dense1,dense2}, her result
to Bob. This, besides turning the OP all-quantum, would be more economical
since Alice can use just one qubit to send Bob two classical bits.
In the two nonlocal channels AQT, after Alice's joint measurement
on the first channel, she only needs to apply local operations on
the state of the particle composing the second quantum channel to
convert one state of the Bell basis into another, as for example,
$\sigma_{z}\left|\psi^{\pm}\right\rangle =\left|\psi^{\mp}\right\rangle $;
$\sigma_{x}\left|\phi^{\pm}\right\rangle =\left|\psi^{\pm}\right\rangle $;
$\sigma_{x}\sigma_{z}\left|\phi^{\pm}\right\rangle =\left|\psi^{\mp}\right\rangle $,
thus encoding the message $\left\{ 00,01,10,11\right\} $ into one
of the four Bell states $\left\{ \left|\psi^{\pm}\right\rangle ,\left|\phi^{\pm}\right\rangle \right\} $,
and then sending her qubit to Bob who, in turn, must read which state
he is receiving. Bob can perform the measurement at least in two ways.
(i) He can perform a nonlocal QND measurement, see Fig.1, on the particles
composing the second quantum channel to read in which of the four
Bell state these particles are in, or (ii) he can perform a CNOT unitary
operation followed by a Hadamard gate on the entangled qubit, leading
the entangled state to one of these four states: $\left|00\right\rangle _{AC}$,
$\left|01\right\rangle _{AC}$, $\left|10\right\rangle _{AC}$, or
$\left|11\right\rangle _{AC}$, depending on which Bell state it was.
Therefore, by performing a projective measurement on the two-particle
system composing the second channel, Bob decodes the Alice's message
encoded into the state of particles $A$ and $C$ and applies one
of the four local unitary operation $\mathbf{1},$ or $\sigma_{z}$,
or $\sigma_{x}$, or $\sigma_{x}\sigma_{z}$ on the state of particle
$B$ to complete the two channels AQT process.

It is interesting to note that superdense coding, a technique used
to send two bits of classical information using only one qubit, is
the \emph{inverse} of quantum teleportation, which sends one qubit
with two classical bits. The present proposal thus combines two quantum
protocols, which one being the inverse of the other, to transmit information
between two remote places using an all-quantum teleportation protocol.

\emph{Security}. In the single channel AQT, Alice sends Bob an entangled
state of particles $A$ and $C$. If an eavesdropper (Eve) has access
to a joint QND measurement device, she can intercept particles $A$
and $C$ to access the Bell states sent by Alice to Bob. Since the
state of particle C was destroyed by the Alice's joint QND measurement,
she cannot extract useful information from the Bell state only. This
is similar to what occurs within the OP, where Eve can intercept the
classical message sent by Alice but cannot decode the message. Regarding
AQT with superdense coding, if Eve intercepts Alice's qubit during
its flight to Bob, what is obtained is just a part of an entangled
state, with no useful information as well. Therefore, AQT is as secure
as the OP.

\emph{Conclusion}. I have proposed an all-quantum protocol for teleporting
unknown quantum states, which can be done in two different ways. In
the first way, the classical channel is abolished altogether and Alice
and Bob share one single quantum channel. It is important to have
in mind that by one single quantum channel is meant just one pair
of entangled particles with no extra-particles. Alice performs nonlocal
QND measurement on the particle to be teleported and on that composing
the nonlocal state shared with Bob. Due to the nonlocal Alice's measurement,
these two particles is now entangled, and since Alice makes a QND
measurement, she is able to perform local unitary operation to convert
one state of the Bell basis into another, and then to encode a two-bit
message to Bob by sending him the two-particle entangled states. Finally,
bob performs another nonlocal QND measurement to know what local operation
he must perform to complete the teleportation process and sends back
Alice one particle of the pair to retrieve the nonlocal channel. The
single channel AQT protocol is thus prompt to be reinitiated. Ideally,
just one single Einstein-Podolsky-Rosen (EPR) pair is consumed to
achieve N runs of single channel AQT, in contrast with the OP that
consumes N EPR pairs besides needing a classical channel. In the second
way, the original protocol for teleporting unknown quantum states
via dual classical and nonlocal channels is modified by replacing
the classical channel by a quantum one, such that Alice and Bob share
two quantum channels. Then, after Alice's nonlocal QND measurement,
she makes a local unitary operation on the particle of the second
nonlocal channel shared with Bob and sends him two bits of information
through a single qubit using the superdense coding protocol, which
is the most economical way to transmit quantum information. By performing
another joint measurement, Bob decodes the information to know which
of four local operation he must apply to complete the two nonlocal
channels AQT.

It is worthwhile to stress that, differently from classical information,
which can be stored, copied and transmitted easily, single channel
AQT, although theoretically simple, may pose some experimental challenges
related to the storage and transmission of the quantum information
by both Alice and Bob. However, AQT, in its two versions presented
here, is by far more economical then the conventional teleportation
protocol. 
\begin{acknowledgments}
We acknowledge financial support from the Brazilian agency CNPq, CAPES
and FAPEG. This work was performed as part of the Brazilian National
Institute of Science and Technology (INCT) for Quantum Information.
I thank Dr. W.B. Cardoso for useful discussion concerning security
of these AQT. 
\end{acknowledgments}

\end{document}